# Electrically Induced, Non-Volatile, Metal Insulator Transition in a Ferroelectric Gated MoS$_2$ Transistor

Zhongyuan Lu[1], Claudy Serrao[1], Asif I. Khan[1], James D. Clarkson[2], Justin C. Wong[1], Ramamoorthy Ramesh[2,3,4], and Sayeef Salahuddin[1,4,*]

[1]Department of Electrical Engineering and Computer Sciences, University of California, Berkeley, California 94720, USA.

[2]Department of Materials Science and Engineering, University of California, Berkeley, California 94720, USA.

[3]Department of Physics, University of California, Berkeley, California 94720, USA.

[4]Lawrence Berkeley National Laboratory, Berkeley, CA 94720, USA

*email: sayeef@berkeley.edu

**Abstract:**

We demonstrate an electrically induced, non-volatile, metal-insulator phase transition in a MoS$_2$ transistor. A single crystalline, epitaxially grown, PbZr$_{0.2}$Ti$_{0.8}$O$_3$ (PZT) was placed in the gate of a field effect transistor made of thin film MoS$_2$. When a gate voltage is applied to this ferroelectric gated transistor, a clear transition from insulator to metal and vice versa is observed. Importantly, when the gate voltage is turned off, the remnant polarization in the ferroelectric can keep the MoS$_2$ in its original phase, thereby providing a non-volatile state. Thus a metallic or insulating phase can be written, erased or retained simply by applying a gate voltage to the transistor.

Due to their unique bandstructure, 2D materials[1–4] could lead to physical phenomena that are otherwise unattainable in conventional semiconductors. For example, the hopping states inside the crystal and the hybridized d-orbitals in the conduction band allow for metal-insulator transition (MIT) and superconductivity in $MoS_2$[5–8], when the charge density reaches a certain threshold. This indicates that if the Fermi-level can be pushed high enough into the conduction band (typically leading to an electron density $>10^{12}/cm^2$), potentially using an electrostatic gate, one should be able to induce phase transition using a gate voltage. However, this has proved to be difficult in practices due to mainly two reasons: (i) the required charge density is high (typically $>10^{12}/cm^2$) and (ii) the 2D transistors are often plagued by the interface states at the semiconductor-oxide surface that greatly reduces the efficacy of the gate. One potential solution to induce such high density is to use ferroelectric materials as the gate oxide and take advantage of the large polarization charge density. For example, single crystalline, perovskite $PbZr_xT_{1-x}O_3$ (PZT) and $BaTiO_3$(BTO) give remnant polarizations of 50~80 $\mu C/cm^2$ [9,10], that is equivalent to a charge density of ($>10^{14}/cm^2$). In addition, due to the remanant polarization of the ferroelectrics, such phase transition can be non-volatile. However, integrating a perovskite ferroelectric with 2D materials in general has proved to be significantly challenging[11–15], due to the interface charge that accumulates on the polar ferroelectric surface. In fact, these interface states completely screen out polarization, leading to the so-called 'anti-hysteresis' observed in almost all experiments. Recent experiments done using polycrystalline, doped $HfO_2$ on $MoS_2$[16–19], within the context of Negative Capacitance Transistors[20] shows ferroelectric polarization induced control of channel charge. In addition, polymer ferroelectric gated 2D channel based transistors have been demonstrated (see, for example, [21] and references therein). Nonetheless, polarization switching induced phase transition has yet to be demonstrated. In this paper, we demonstrate such an electrically induced, non-volatile phase transition, using epitaxially grown, single crystalline PZT as a gate oxide in a MoS$_2$ Field Effect Transistor.

MoS$_2$ flakes were mechanically exfoliated from bulk crystals and transferred on to 270 nm SiO$_2$/Si substrates. Optical microscopy was used to locate the thin flakes (Figure 1a), and atomic force microscopy (AFM) was used to measure their precise thicknesses (2.1 nm corresponds to a trilayer MoS$_2$ flake) (Figure 1b). A mesa region was defined as the transistor's channel and the remaining parts were etched away by XeF$_2$ gas[22]. Figure 1c shows the schematic structure of the designed top-gated MoS$_2$ transistor. Au(50 nm)/Ti(10 nm) source/drain electrodes were evaporated after e-beam lithography patterning. The channel length $L_{total}$ is 8 μm, and the width $W$ is 5 μm. 20 nm of Al$_2$O$_3$ were deposited over the MoS$_2$ channel via 200 °C thermal atomic layer deposition (ALD) to form the gate dielectric using 1 nm of evaporated SiO$_x$ as the nucleation layer[23]. Finally, the top gate electrode was defined ($L_g$ = 5 um) and patterned with Au(50 nm)/Ti(10 nm). A 40 μm × 40 μm dielectric capacitor was added in parallel, with pads connected to the source and gate separately (Figure 1d). The rationale for using this parallel capacitor, that increases the total capacitance seen from the gate of the transistor to the ground, will be discussed later.

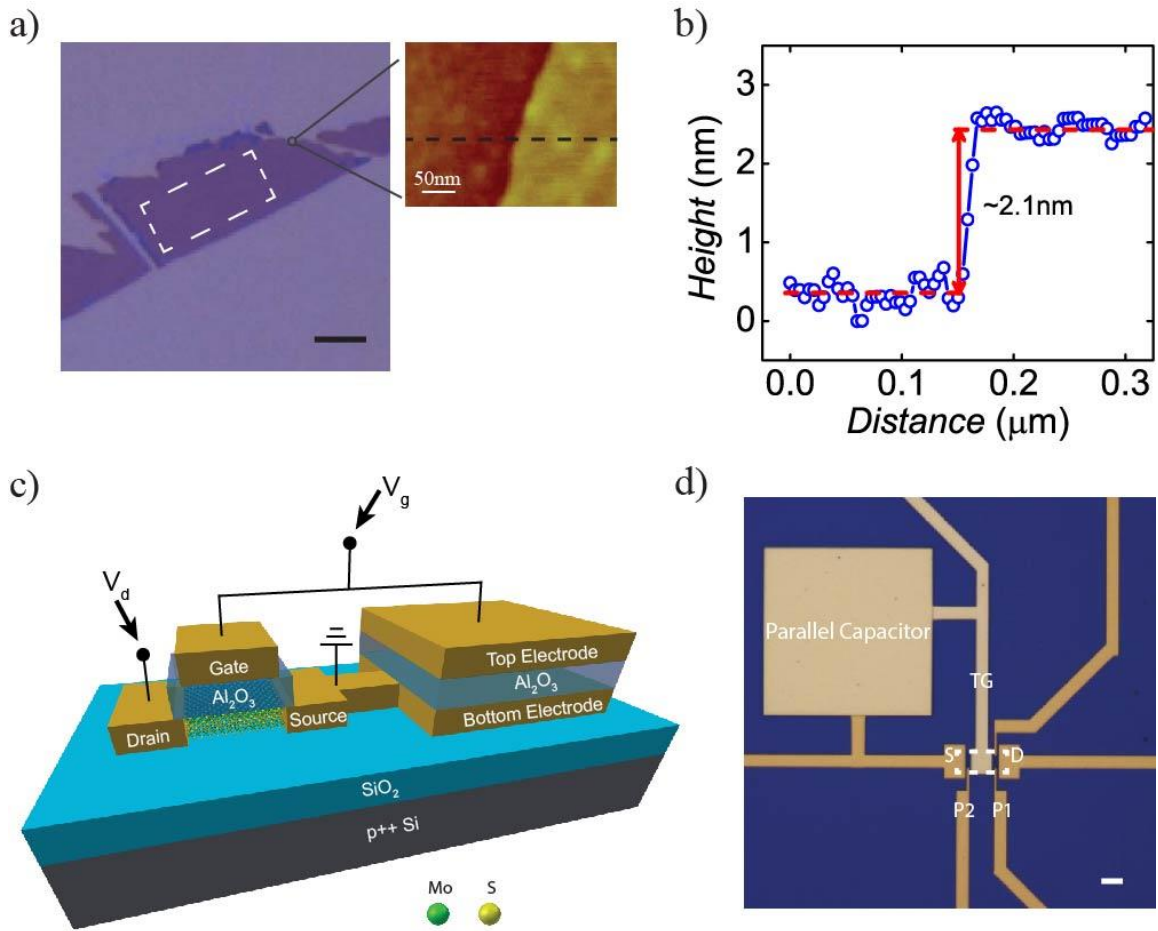

Figure 1. a) The optical image and AFM topography of the MoS$_2$ flake. The region enclosed by dashed lines is the remaining channel area after dry etching. The scale bar is 5 μm. b) The step height value measured by AFM is ~2.1 nm. c) A 3D schematic view of the MoS$_2$ top gate transistor with the parallel capacitor. d) Optical image of fabricated top gate transistor. The channel is outlined with dashed lines, and the scale bar is 5 μm.

The transistor's electrical properties were characterized by an Agilent B1500A device analyzer in a high vacuum environment (~2×10$^{-6}$ Torr) at room temperature (300 K). The transistor shows an ON-OFF ratio (>10$^6$) and a subthreshold swing of ~166 mV/decade (Fig 2a). The clockwise hysteresis loop in the transfer curves originates from interface trapping states[24,25]. The ON current is ~28.5 μA (Figure 2b), which is consistent to that observed typically[26,27]. The total capacitance of the entire system (i.e. the transistor combined with the parallel capacitor) was measured as 4.4 pF (Figure 2c). This value did not vary with gate bias at low frequency (100 kHz), implying that the parallel capacitor dominates the total capacitance. The dielectric constant of the gate stack was calculated to be 6.22 (including $Al_2O_3$ and 1 nm $SiO_x$ nucleation layer).

Four-terminal measurement was performed while sweeping the gate voltage in order to exclude contact resistance at the source/drain. The field effect mobility $\mu_{FE}$ was computed to be ~14.5 $cm^2\ V^{-1}\ s^{-1}$ using the equation

$$\mu_{FE} = \left(\frac{L_{12}}{W}\right) \times \frac{dG/dV_g}{c_g} \tag{1}$$

where $c_g$ = 0.275 µF/cm² is the gate capacitance per unit area, $G$ is the channel conductance (Figure S2a), and $L_{12}$ = 6 µm and $W$ = 5 µm represent the probe separation distance and the channel width respectively.

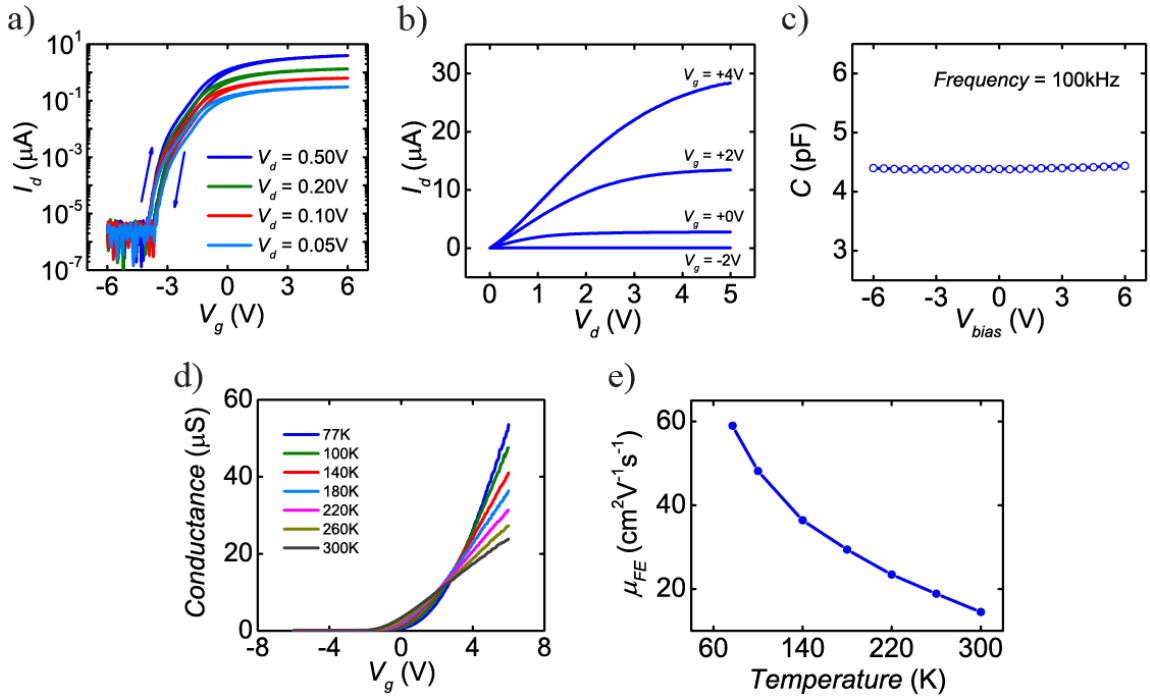

Figure 2. a) Transfer curves of the top gate transistor. b) Output characteristics. c) The device's total capacitance measured at 100 kHz. d) Temperature-dependent conductance measurement. e) Field-effect mobility values at different temperatures.

Figure 2d shows the dependence of channel conductance on gate voltage at different temperatures. There is a well-defined crossover point at $V_g$ = 2.61 V, which represents the transition from an insulating phase to a metallic phase. In the low gate voltage region, MoS$_2$ is in the insulating phase in which electrons are primarily localized by impurities or localized states and transport through variable range hopping[28]. As temperature increases, thermal excitation enhances the hopping transport, resulting in higher channel conductance. With large enough gate voltage, the carrier density becomes high enough to result in strong Coulomb interactions between electrons to screen the localization states—phonon scattering becomes the primary scattering mechanism, and the phase becomes metallic. Hence at lower temperatures, the phonon energies decrease, and the conductance increases. The critical carrier density at which MIT occurs for this trilayer MoS$_2$ flake is $n_c \approx 5.64 \times 10^{12}$ cm$^{-2}$, which was calculated from

$$n_c = c_g(V_g - V_t)/q \qquad (2)$$

where $V_g$ = 2.61 V is the gate voltage at the crossover point, $V_t$ = -0.67 V is the threshold voltage (Figure S2), and $q = 1.6 \times 10^{19}$ C represents the charge of an electron. This carrier density value is similar to that of previous reports[8,29]. Figure 2e shows the temperature-dependence of electron mobility over a range of temperatures in which phonon scattering dominates. Thus, as the device is cooled, thermal vibrations inside MoS$_2$ become weaker, resulting in the mobility to gradually increase to its maximum value of 59 cm$^2$ V$^{-1}$ s$^{-1}$ at 77 K. The mobility has the temperature dependence $\mu \propto T^{-\gamma}$ where $\gamma$ is ~1.8 at 300 K and ~0.8 at 77 K (Figure S2b). This variation in $\gamma$ could be attributed to different phonon contributions (e.g. optical at high temperatures and acoustic at low temperatures) as well as variable-range hopping.

Now that it is established that the material quality of our transistors allows for metal-insulator phase transition and the electron density necessary to induce such a phase transition is determined from the baseline transistor, we have designed a single crystalline, ferroelectric (FE) PZT capacitor that is then connected to the gate of the baseline transistor (Figure 3a, [30]). The PZT film's electrical characteristics are shown in Figure 3c and Figure S1. The film shows a high remnant polarization of ~65 µC/cm$^2$, and the completely closed hysteresis loop indicates insignificant leakage. The capacitance of the fabricated 30 µm × 30 µm PZT capacitor (Figure S1b) varied between 5.27 pF and 8.32 pF (Figure S1d), corresponding to its nonlinear polarization properties. Typically, this capacitance will be much larger than a transistor capacitance. This means that when the ferroelectric capacitor is connected to the gate of the MoS$_2$ transistor (see Fig. 3b), almost all the voltage will drop across the transistor, making it impossible to switch the ferroelectric and thereby achieve electronically tunable metal insulator transition. We have identified this to be a critical issue in other experimental works on ferroelectric gate TMD transistors where a metal insulator phase transition was not observed. To circumvent this effect in our work, we have designed the additional parallel capacitor, so that the capacitance of the combined transistor and parallel capacitor is comparable to that of the PZT capacitor.

The transfer characteristics of the combined system (with the external FE capacitor connected) are shown in Figure 3d. The counterclockwise hysteresis loops correspond to the ferroelectric switching. The loops at different drain voltages have the same window size and start/end points, which means that the polarization switching induced electrostatic doping mechanism is stable.

Shifts in threshold voltage follow the switches of the external FE capacitor (Figure 3e). When sweeping in the forward direction, the threshold voltage is shifted positive relative to that of the baseline transistor, implying that the ferroelectric polarization is pointing out of the transistor and depleting the channel of electron carriers. At the end of the forward sweep (i.e. after reaching the coercive voltage), the ferroelectric polarization switches to point into the transistor, enhancing the electron carrier density in the channel. Thus, when sweeping in the reverse direction, the threshold voltage is shifted negative relative to that of the baseline transistor. Interestingly, when the baseline transistor is probed in isolation, it shows a clock-wise hysteresis; but the same transistor with a ferroelectric capacitor at the gate shows an anti-clockwise hysteresis. This shows that the large polarization of the ferroelectric can completely dominate the surface charges that come from interfacial traps.

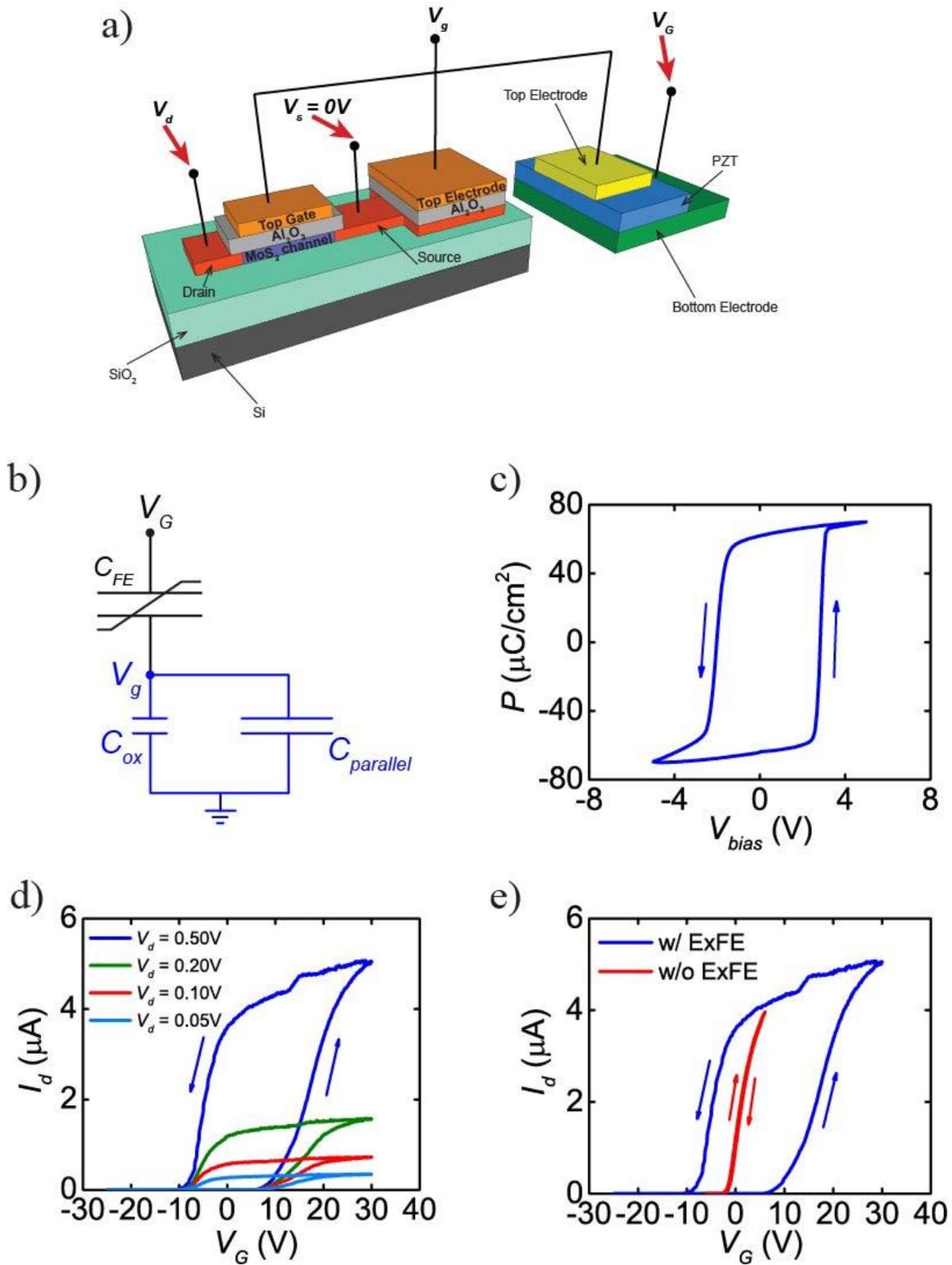

Figure 3. a) 3D schematic of the transistor and parallel capacitor connected with an external PZT capacitor. b) Equivalent circuit model of the overall system. c) Polarization versus voltage curve for the PZT capacitor. d)

Transfer curves for different drain voltages when the system is connected to an external PZT capacitor. e) Transfer characteristics of the MoS$_2$/Al$_2$O$_3$ transistor with and without the external PZT capacitor connection ($V_d$ = 0.5 V).

The effective threshold voltage shift $\Delta V_{\text{geff}}$ in the MoS$_2$ transistor is 10.2 V, derived by

$$\Delta V_{\text{geff}} = \Delta V_G \left( \frac{C_{\text{FE}}}{C_{\text{FE}} + C_{\text{DE}}} \right) \quad (3)$$

in which $\Delta V_G$ is the width of the hysteresis loop in Figure 3d. From here, the doping density induced by the ferroelectric switching is $\Delta n_{\text{doped}} = \frac{1}{2}\left( \frac{c_{\text{ox}} \Delta V_{\text{geff}}}{q} \right) \approx 7.8 \times 10^{12}$ cm$^{-2}$, which is larger than the critical charge density $n_c = 5.64 \times 10^{12}$ cm$^{-2}$ needed to trigger the phase transition. This indicates that it is feasible to control the MIT with a series FE capacitor. The electrical characteristics of PZT at different temperatures are depicted in Figure 4a. Notice that the coercive voltage increases as the temperature decreases while the remnant polarization remains virtually unchanged[31]. Since the remnant polarization is approximately constant, the electron doping density at different temperatures can also be regarded as unchanged. Figure 4b shows the temperature-dependent changes in conductance for our combined system. Notice that the counterclockwise hysteresis loops consistently remain intact, and there is a crossover point in the forward direction. This indicates that after a certain amount of positive voltage is applied, an insulator-to-metal phase transition should happen. To test this hypothesis, in Fig. 4(c), we have plotted conductance as a function of temperature for $V_g$ = 13 V for the two branches shown in Fig. 4(b). Note that for smaller $V_g$'s, the conductance values in the forward sweep branch is too small compared to the noise floor to provide a meaningful trend. Nonetheless, at $V_g$ = 13 V, one can clearly see that two distinct phases are seen for two different directions of the sweep. In other words, a non-volatile state is achieved where the MoS$_2$ transistor can be placed in either the metallic or insulating phase by applying a gate voltage and kept there indefinitely until a RESET voltage is applied.

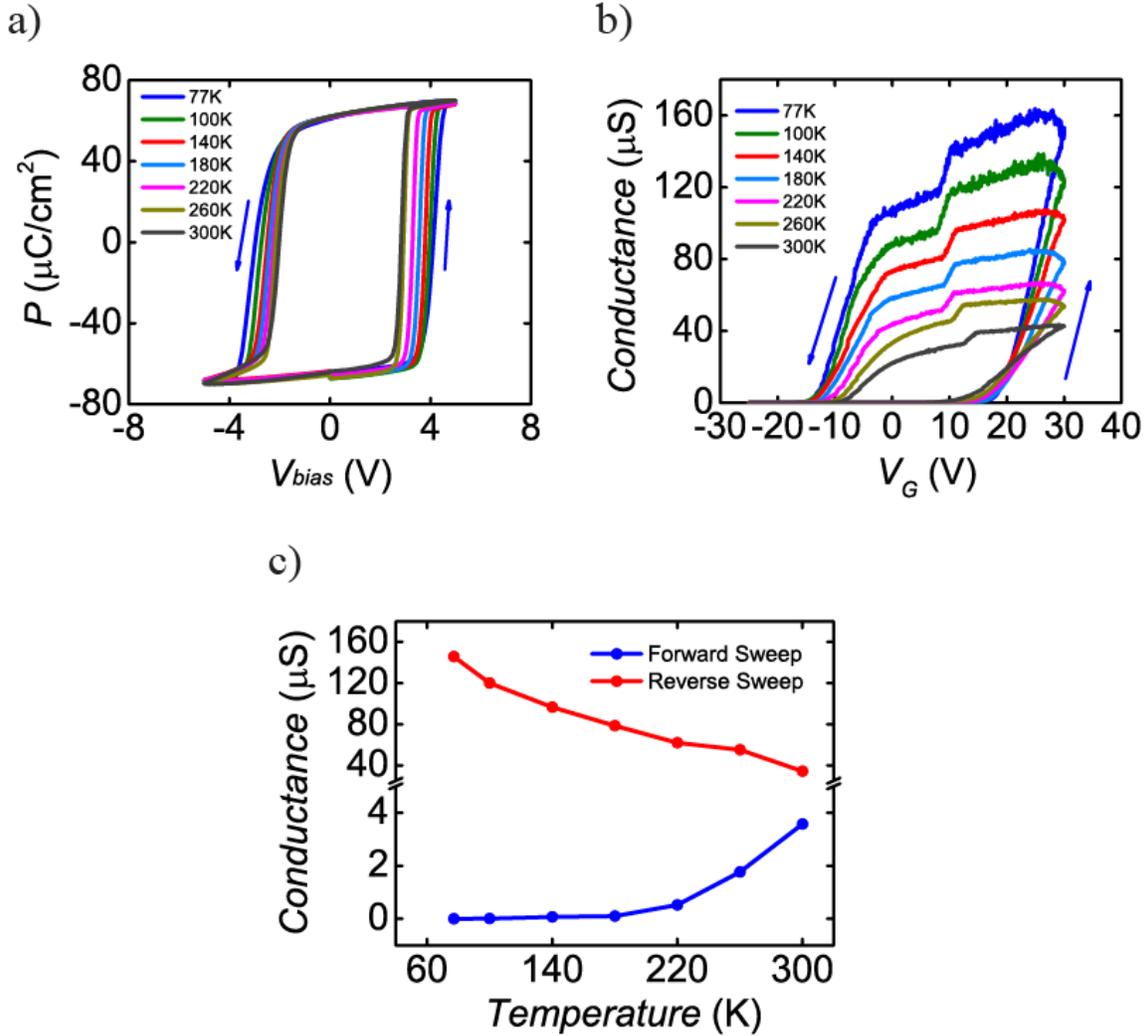

Figure 4. a) Hysteresis loop of the PZT capacitor at different temperatures. b) Conductance versus gate voltage at different temperatures when the external FE capacitor is connected. c) Conductance versus temperature at the same gate voltage (+13 V) when the temperature is swept in the forward and reverse directions.

In summary, we have demonstrated that it is possible to electrically induce a non-volatile metal insulator phase transition in $MoS_2$ transistors. This is achieved by placing a high remnant polarization ferroelectric capacitor at the gate. As the polarization switches, it induces the metal insulator transition. Importantly, as the gate voltage is then put back to zero, the remnant polarization is strong enough to retain the phase of the $MoS_2$. Our results could lead to new opportunities for non-volatile memory applications. In addition, such non-volatile metal insulator phase transition could lead to an electrically switchable, non-volatile, transition to superconductivity in 2D materials[32,33].

# Supplementary Information

Electrically Induced, Non-Volatile, Metal Insulator Transition in a Ferroelectric Gated MoS$_2$ Transistor

Zhongyuan Lu[1], Claudy Serrao[1], Asif I. Khan[1], James D. Clarkson[2], Justin C. Wong[1], Ramamoorthy Ramesh[2,3,4], and Sayeef Salahuddin[1,4,*]

[1]*Department of Electrical Engineering and Computer Sciences, University of California, Berkeley, California 94720, USA.*

[2]*Department of Materials Science and Engineering, University of California, Berkeley, California 94720, USA.*

[3]*Department of Physics, University of California, Berkeley, California 94720, USA.*

[4]*Lawrence Berkeley National Laboratory, Berkeley, CA 94720, USA*

*sayeef@berkeley.edu

# Sample Preparation

The MoS$_2$ bulk crystal was purchased from SPI and was mechanically exfoliated using adhesive tape. The SiO$_2$/Si wafer was produced by Silicon Valley Microelectronics (SVM). PZT films of thickness ~100 nm were grown via pulsed laser deposition (PLD) using a KrF excimer laser (wavelength = 248 nm) and a thin conductive SrRuO$_3$ (SRO) buffer layer (25 nm) as the back electrode. The grown PZT and SRO films are both confirmed to be single crystalline with those sharp peaks in the X-ray diffraction (XRD) spectrum of the PZT/SRO/STO stack (Figure S1a). A platinum (Pt) electrode was sputtered onto PZT (Figure S1b) and coupled with the SRO conductive layer to form an FE capacitor (Figure S1c). C-V measurements were performed at 100 kHz with an Agilent B1500A. The admittance angles (Figure S1d) were nearly 90° at high voltages, indicating excellent insulating properties of the grown PZT film.

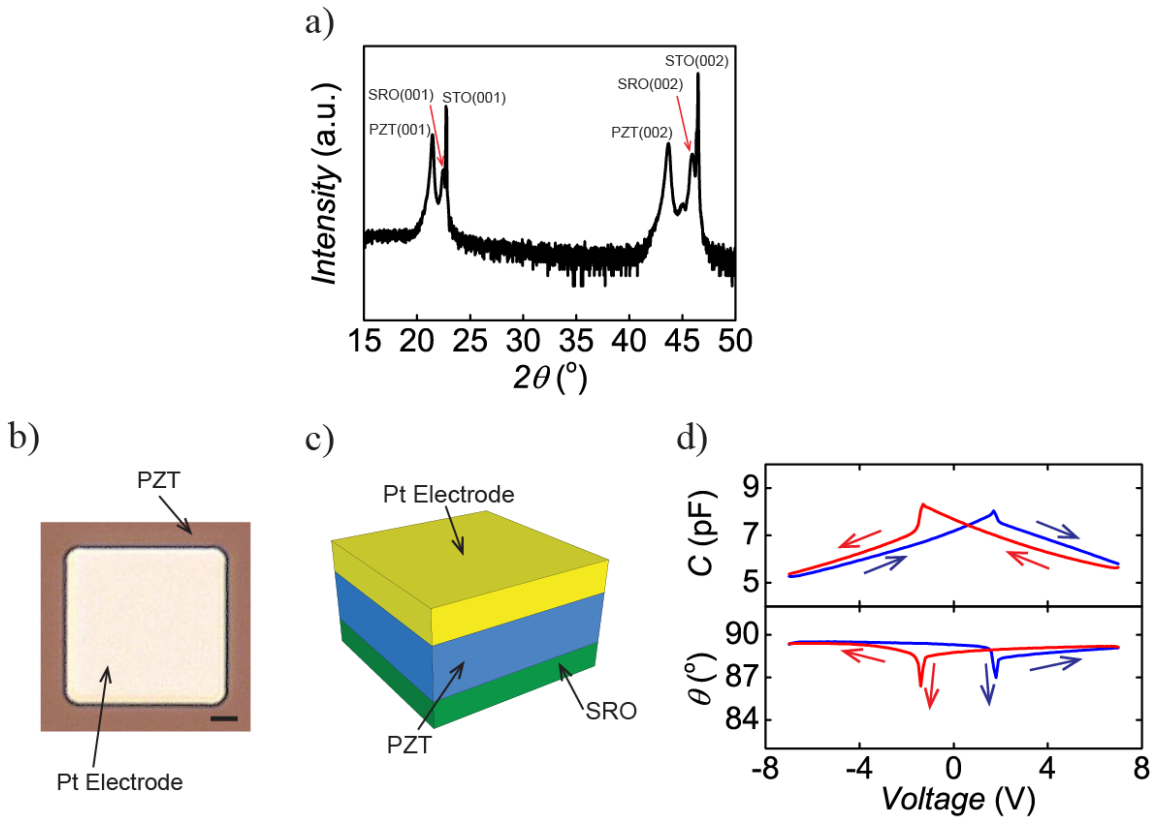

Figure S1. Physical and electrical properties of the PZT capacitor. a) X-ray diffraction (XRD) spectrum of the PZT/SRO/STO heterostructure. b) Optical image of the PZT metal-insulator-metal capacitor. The scale bar is 5 µm. c) 3D schematic of the FE capacitor. d) The FE capacitor's capacitance and admittance angle $\theta$ versus voltage.

## Additional Electrical Characteristics

The conductance versus gate voltage for the MoS$_2$ top gate transistor is shown in Figure S2a. Here the threshold voltage $V_t$ was extracted as -0.67 V. Based on the measured drain current $I_d$ and the voltage difference $V_{12}$ between the 2 probes on the channel, the channel conductance $G$ is:

$$G = \frac{I_d}{V_{12}}$$

The slope 3.32 µS/V in the linear region of Figure S2a represents $dG/dV_g$, which is used in equation (2) for the field effect mobility $\mu_{FE}$ calculation.

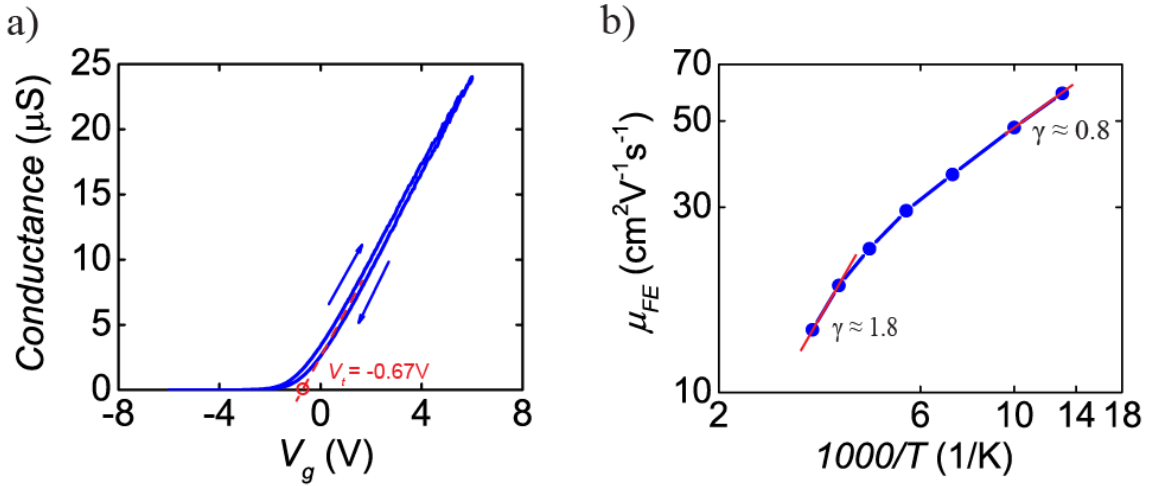

Figure S2. a) Conductance versus gate voltage. The threshold voltage is marked as -0.67 V. b) Log-log plot of field-effect mobility $\mu_{FE}$ versus inverse temperature 1000/T.

The mobility versus temperature is plotted in Figure S2b. The slope here is the damping factor $\gamma$. At room temperature, optical phonon scattering dominates the mobility damping. Due to the quenching of homopolar phonon modes by the encapsulated structure, monolayer $MoS_2$ is expected to have a $\gamma$ of ~1.52[S1]. Our result $\gamma$ ~ 1.8 is similar to that of other reports about multilayer films[S2,S3], implying that thin film $MoS_2$ still behaves much like a 2D system. At 77 K, most optical phonons are frozen, allowing acoustic phonons ($\gamma = 1$)[S1] and variable-range hopping ($\gamma = 1/3$)[S4] to play leading roles in limiting the mobility and contributing to the final result: $\gamma$ ~ 0.8.